\documentclass[11pt,twoside]{article}

\usepackage{asp2006}
\usepackage{epsf}
\usepackage{psfig}
\usepackage{lscape}

\begin{document}

\newcommand{\msun}{\mbox{${\cal M}_\odot$}}
\newcommand{\lsun}{\mbox{${\cal L}_\odot$}}
\newcommand{\kms}{\mbox{km s$^{-1}$}}
\newcommand{\HI}{\mbox{H\,{\sc i}}}
\newcommand{\mhi}{\mbox{$M_{\rm HI}$}}
\newcommand{\HII}{\mbox{H\,{\sc ii}}}
\newcommand{\am}[2]{$#1'\,\hspace{-1.7mm}.\hspace{.0mm}#2$}
\newcommand{\as}[2]{$#1''\,\hspace{-1.7mm}.\hspace{.0mm}#2$}
\newcommand{\ad}[2]{$#1^{\circ}\,\hspace{-1.7mm}.\hspace{.0mm}#2$}
\newcommand{\lsim}{~\rlap{$<$}{\lower 1.0ex\hbox{$\sim$}}}
\newcommand{\gsim}{~\rlap{$>$}{\lower 1.0ex\hbox{$\sim$}}}
\newcommand{\dark}{$M_{HI}/L_{B}$}
\newcommand{\nan}{Nan\c{c}ay}

\title{VLA Observations of H\,{\bf{\sc i}} in the Circumstellar 
Envelopes \\ of AGB Stars}
\author{Lynn D.~Matthews and Mark J.~Reid}
\affil{Harvard-Smithsonian Center for Astrophysics, Cambridge MA, USA}

\begin{abstract}
We present the results of a VLA search for H\,{\sc i} emission 
in the circumstellar envelopes of five nearby AGB stars: RS~Cnc, 
IRC+10216, EP~Aqr, R~Cas, and R~Aqr. We have detected emission 
coincident in both position and velocity with RS~Cnc, implying that 
the emission arises from its extended envelope. For R~Cas, we 
detected weak ($5\,\sigma$) emission that peaks at the stellar 
systemic velocity and overlaps with the location of its
circumstellar dust shell and thus is probably related to the star.
Toward IRC+10216 and EP~Aqr, we detected multiple, arcminute-scale 
H\,{\sc i} emission features at velocities consistent with the 
circumstellar envelopes, but spatially offset from the stellar 
positions; in these cases we cannot determine unambiguously if the 
emission is associated with the stars.
In the case of IRC+10216, we were unable to confirm the detection 
of H\,{\sc i} in absorption against the cosmic background previously 
reported by Le~Bertre \& G\'erard.
We detected our fifth target, R~Aqr (a symbiotic binary), in the 
1.4~GHz continuum.
\end{abstract}

\section{Summary}

Recent single-dish surveys have established that neutral atomic 
hydrogen (H\,{\sc i}) is common in the circumstellar envelopes of 
evolved, low-to-intermediate mass stars undergoing mass-loss 
(G\'erard \& Le~Bertre 2006 and references therein). The large 
extents of the envelopes detected in H\,{\sc i} 
(up to $\sim$2~pc) imply that H\,{\sc i} probes different regions 
of the envelope than CO or other molecular tracers and thus can 
trace mass-loss over very large time-scales --- up to $\sim10^{5}$\,yr. 
Studies of the 21-cm emission from circumstellar envelopes can 
therefore provide important new constraints on atmospheric models 
of AGB stars, the physical conditions in their extended envelopes, 
and on the rates, timescales, and geometries of their mass-loss.

We have recently undertaken a VLA H\,{\sc i} imaging survey of five 
nearby AGB stars: RS~Cnc, IRC+10216, EP~Aqr, R~Cas, and R~Aqr. 
H\,{\sc i} detections of four of these targets 
(RS~Cnc, EP~Aqr, and R~Cas in emission and IRC+10216 in absorption) 
have been published previously based on single-dish observations 
(Le~Bertre \& G\'erard 2001, 2004; G\'erard \& Le~Bertre 2003, 2006).

We have confirmed the presence of H\,{\sc i} emission coincident in 
position and velocity with the semi-regular variable RS~Cnc, implying 
that the emission is indeed associated with its circumstellar envelope 
(Figure 1). We estimate a total H\,{\sc i} mass for this material of
$M_{\rm HI}\approx 1.5\times10^{-3}M_{\odot}$ (for $d$ = 122~pc).
The morphology of the emission suggests that a component of the 
mass-loss 
is highly asymmetric. From the H\,{\sc i} data 
we derive a recent mass-loss rate of 
${\dot M}=1.7\times10^{-7}M_{\odot}$~yr$^{-1}$, 
comparable to previous estimates based on CO observations.

\begin{figure}[!ht]
\plotfiddle{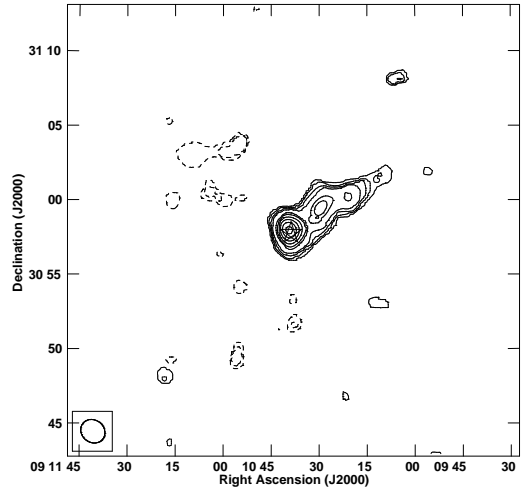}{75mm}{0}{35}{35}{-110}{-5}
\vspace{-1.5cm}
\caption{VLA \HI\ total intensity map of a 30$'$ region around RS~Cnc. 
The contour levels are 
(--16,\,--8, 8, 16, 24\,...\,86)$\times$1.25~Jy beam$^{-1}$ m s$^{-1}$. 
A star symbol marks the optical position of the star.}
\end{figure}

For the Mira variable R~Cas we have detected weak (5\,$\sigma$) 
emission centered at the systemic velocity of the star. The morphology 
of the emission is consistent with a partial shell-like structure with 
a radius $r\sim100''$. This structure overlaps with the dust shell 
previously detected by Young et al.~(1993a,b), and we estimate for it 
an H\,{\sc i} mass of 
$M_{\rm HI}\approx 5.3\times10^{-4}M_{\odot}$, 
assuming $d$ = 160~pc.

Toward two other targets (the carbon star IRC$+$10216 and the
semi-regular variable EP~Aqr) we have detected multiple 
arcminute-scale H\,{\sc i} emission features at velocities 
consistent with the circumstellar envelope, but spatially offset from 
the position of the stars.  In these cases we cannot determine 
unambiguously whether the emission arises from material within the 
circumstellar envelope or, instead, from the chance superposition 
of Galactic H\,{\sc i} clouds along the line of sight.

We detected our fifth target, R~Aqr (a symbiotic star with a hot
companion), in the 1.4~GHz continuum with a flux density
$F_{\rm 1.4GHz}=18.8\pm0.7$ mJy. R~Aqr is a well-known radio source,
and the continuum emission likely arises primarily from free-free 
emission from an ionized circumbinary envelope.  However, we did not 
detect any neutral hydrogen associated with this system.

A more extensive discussion of these results can be found in Matthews 
\& Reid (2007).

\end{document}